\documentclass{elsart}
\usepackage{graphicx}
\usepackage{amsmath}
\usepackage{amssymb}
\usepackage{amsfonts}
\usepackage{psfrag}
\usepackage{epsfig}

\newcommand{\T}{\mathbb{T}}

\newtheorem{lemma}{Lemma}

\newcommand{\proof}{\noindent {\bf Proof:}~~}

\begin{document}

\begin{frontmatter}

\title{Desynchronization transitions in nonlinearly coupled phase oscillators}

\author{Oleksandr Burylko}
\address{Institute of Mathematics, National Academy of Sciences of
Ukraine, Tereshchenkivska Str. 3, 01601 Kyiv, Ukraine}
\author{Arkady Pikovsky}
\address{Department of Physics and Astronomy, Potsdam University ,
Karl--Liebknecht-Str. 24/25, D14476 Potsdam--Golm, Germany}

\date{\today}



\begin{abstract}
We consider the nonlinear extension of the Kuramoto model of globally
coupled phase oscillators where the phase shift in the coupling function
depends  on the order parameter. A bifurcation analysis of the 
transition from fully synchronous state
to partial synchrony is performed. We demonstrate that for small ensembles
it is typically mediated by stable cluster states, that disappear with creation of heteroclinic cycles,
while for a larger number of oscillators a direct transition from full synchrony to
a periodic or quasiperiodic regime occurs.
\end{abstract}
\today
\end{frontmatter}

\section{Introduction}

A model of coupled limit cycle oscillators explains
a variety of natural phenomena in various fields of science. The applications 
range from the description of the collective dynamics of 
Josephson junctions \cite{Wiesenfeld-Swift-95}, lasers \cite{Glova-03}, and 
electrochemical oscillators \cite{Kiss-Zhai-Hudson-02a} to
neuronal populations \cite{Golomb-Hansel-Mato-01}, etc.
Very often, when the oscillator network is not too sparse, it can be 
approximately considered as fully connected, or globally coupled.

Ensembles of weakly interacting units are successfully treated within the 
framework of phase approximation 
\cite{Kuramoto-75,Kuramoto-84,Daido-96}.
Most popular is the Kuramoto model of sine-coupled phase oscillators, or its extension, 
the Kuramoto-Saka\-gu\-chi model \cite{Sakaguchi-Kuramoto-86}.
This model explains self-synchroni\-za\-tion and appearance of a collective mode 
(mean field) in
an ensemble of generally non-identical elements; the transition to synchrony occurs 
at a certain critical value of the coupling constant that is roughly proportional 
to the width of the distribution of natural 
frequencies \cite{Kuramoto-75,Kuramoto-84,Pikovsky-Rosenblum-Kurths-01,Strogatz-00}.

An extension of the Kuramoto model for the case of nonlinear coupling has been suggested
in our recent publications \cite{Rosenblum-Pikovsky-07,Pikovsky-Rosenblum-09}, see also
\cite{Filatrella-Pedersen-Wiesenfeld-07,Giannuzzi_et_al-07}.
Nonlinearity in this context means that the effect of the collective mode on an individual unit
depends on the amplitude of this mode, so that, e.g., the interaction of the field and of a 
unit can be attractive for a weak field and repulsive for a strong one. 
Formally, this is represented 
by the dependence of the parameters of the Kuramoto-Sakaguchi model (the coupling strength and 
the phase shift) on the mean field amplitude.
The model exhibits nontrivial effects like a 
destruction of a completely synchronous state and appearance of partial synchrony in an 
ensemble of identical units. Moreover, in this setup the frequencies of the collective mode 
and of oscillators can be different and incommensurate.
 
An analytical description of the dynamics of oscillator ensembles remains an important 
and challenging problem.
A seminal work in this direction is that of Watanabe 
and Strogatz (WS) \cite{Watanabe-Strogatz-93,Watanabe-Strogatz-94}.
The WS theory is a powerful tool that provides a nearly full dynamical description 
of ensembles of \textit{identical} oscillators, sine-coupled to a common external 
force. In particular, this force can be the mean field of the population, so that 
for the case of identical units the WS theory almost completely describes the 
Kuramoto-Sakaguchi and the nonlinear models (see \cite{Pikovsky-Rosenblum-09}).
This description is given in terms of three collective (macroscopic)
variables, hereafter called the WS variables, plus constants of motion. 
The collective variables obey $3$ WS equations (see~\cite{Pikovsky-Rosenblum-11});
thus, the dynamics of an ensemble of identical elements is effectively 
$3$-dimensional. However, the WS theory has one drawback: it cannot describe 
certain cluster states, i.e. regimes where the oscillators build identical groups. In this paper we
complement the WS theory by performing a direct bifurcation analysis of the dynamical phase equations
of the model of nonlinearly coupled phase oscillators. We will especially emphasis on cluster states and
their bifurcations, in particular on the heteroclinic cycles (see \cite{DynSys-special-issue-10} for a recent review
of robust heteroclinic cycles)
that can be hardly treated within the WS approach. We will see that the role of clusters is mostly important
for small ensembles. Because of identity of the oscillators, the system possesses a permutation symmetry,
so we employ the corresponding bifurcation approach (see, e.g., \cite{Golubitsky-Stewart-02,Ashwin-etal-06}).

The paper is organized as follows. We introduce the basic model in Section~\ref{sec:mncpo}. 
Then in Section~\ref{sec:gasb} we discuss general properties of bifurcations, possible attractors and their
interpretation as different synchronization patterns. In Section~\ref{sec:ncowqpn} we present bifurcation diagrams
for a model of nonlinearly coupled oscillators with quadratic nonlinearity~\cite{Pikovsky-Rosenblum-09}. 
In Conclusion a relation to the WS theory is discussed.

\section{Model of nonlinearly coupled phase oscillators}
\label{sec:mncpo}
We consider an ensemble of $N$ limit cycle oscillators, described by their phases $\theta_i\in[0,2\pi)$, $i=1,\ldots,N$. They are assumed to interact globally, via the complex mean field 
\begin{equation}
r e^{i \psi}=\frac{1}{N}\sum_{j=1}^N e^{i \theta_j} \; ,
\label{eq:cmf}
\end{equation}
having amplitude $r$ and phase $\psi$:
\begin{equation}
\dot\theta_i=\omega_i+G(r,\psi,\theta_i)\;.
\label{eq:gcpl}
\end{equation}
Here $\omega_i$ are
natural frequencies of the oscillators and $G$ is the coupling function.
Different popular models correspond to different choices of coupling function $G$. 
The case $G(r,\psi,\theta_i)=rK\text{Im}(e^{i(\psi-\theta)})$ 
corresponds to the famous Kuramoto 
model~\cite{Kuramoto-84}, while the choice 
$G(r,\psi,\theta_i)=rK\text{Im}(e^{-i\alpha} e^{i(\psi-\theta)})$ yields the 
Kuramoto-Sakaguchi model~\cite{Sakaguchi-Kuramoto-86}.

In this paper we focus on a coupling function 
that nonlinearly depends on the amplitude
of the mean field $r$ and on a set of parameters $\beta$:
\begin{equation}
G(r,\psi,\theta_i)=rK(r,\beta)\sin(\psi-\theta+\alpha(r,\beta))\;.
\label{eq:nop}
\end{equation} 
This model has been introduced in ~\cite{Rosenblum-Pikovsky-07} and studied in the thermodynamic 
limit $N\to\infty$ in~\cite{Pikovsky-Rosenblum-09}. 
In this paper we focus on the properties of small ensembles of nonlinearly coupled 
oscillators, restricting our analysis to the case of identical oscillators $\omega_i=\omega$
and of phase nonlinearity only $K(r,\beta)=1$. The latter restriction is not very important, as
the cases where $K(r,\beta)$ can change sign are in fact trivial. Substituting (\ref{eq:nop}) in
 (\ref{eq:gcpl}) we obtain an equivalent formulation of the ensemble dynamics
\begin{equation}
\dot\theta_i=\omega+\frac{1}{N}\sum_{j=1}^N\sin(\theta_j-\theta_i+\alpha(r,\beta))\;.
\label{eq:gcpl1}
\end{equation}

To exploit the phase-shift symmetry of this system one can
describe the system dynamics in terms of the phase differences
\begin{equation}\label{substitution}
\varphi_i=\theta_1-\theta_{i+1},\quad i=1,\dots,N-1,
\end{equation}
thus reducing this $N$-dimensional system to the
$(N-1)$-dimensional system
\begin{equation}
\label{PhaseDifferences}
\begin{aligned}
\dot\varphi_i&= - \frac{1}{N} \left[
\sum\limits_{j=1, j\not = i}\limits^{N-1}
\sin(\varphi_i-\varphi_j+\alpha(r,\beta))+ \right.
\\
& \left.+\sin(\varphi_i+\alpha(r,\beta))
+\sum\limits_{j=1}\limits^{N-1} \sin(\varphi_j-\alpha(r,\beta))
\right]\;.
\end{aligned}
\end{equation}
One can check that order
parameter $r$ can be written in phase differences as
\begin{equation}\label{radius}
r=\frac{1}{N}\sqrt{N+2\sum_{i,j=1,i\not
=j}^{N-1}(\cos(\varphi_j)+\cos(\varphi_i-\varphi_j))}\;.
\end{equation}
 Below we
will discuss synchronization transitions in the system
studying invariant manifolds, fixed points, cycles, heteroclinic
cycles and their bifurcations for the system in phase differences
(\ref{PhaseDifferences}).

Before proceeding to the analysis, we mention that
system (\ref{eq:gcpl1}) possesses
symmetries given by all permutations of the oscillators
\cite{Ashwin92}. Due to identity of the oscillators, the main dynamical regimes
appear as invariant sets of the system:\\
1) Completely synchronous solution, where all the oscillators are in the same state:
$$
\mathcal{O}=\{(\theta_1,\cdots,\theta_N)~:~\theta_1=\theta_2=\dots=\theta_N\};
$$
3) Completely asynchronous solution
\begin{equation}\label{Manifold}
\mathcal{M}=\left\{ (\theta_1,..., \theta_N): \ \sum\limits_{j=1}^N
e^{i \theta_j}=0 \right\}.
\end{equation}
The set $\mathcal{M}$ is a union of invariant manifolds of dimension
$N-2$ for $N\geq 3$ \cite{Ashwin-etal-08}, it corresponds to the case of
vanishing order parameter $r=0$.\\
2) Cluster states, where groups of oscillators have identical phases.
A general $n$-cluster state can be written as (up to permutation of indices)
\begin{equation}\label{cluster}
\mathcal{P}_n=\left\{ (\theta_1,..., \theta_N): \theta_1=\cdots=\theta_{p_1}\,;\theta_{p_1+1}=
\cdots=\theta_{p_1+p_2}\,;\cdots\,;
\,\theta_{\sum_1^{n-1}p_j+1}=\cdots=\theta_N\right\},
\end{equation}
 where $p_1+p_2+\cdots+p_n=N$. We will be mainly interested in 2-cluster states (we will see that only such states appear as stationary solutions)
\begin{equation}\label{2cluster}
\mathcal{P}_{2}=\left\{ (\theta_1,..., \theta_N): \theta_1=\cdots=\theta_{p}\,;\theta_{p+1}=\cdots=\theta_{N}\right\}\;,
\end{equation}
characterized by the partition $(p:N-p)$.


\section{General analysis of synchronization and bifurcations}
\label{sec:gasb}
In this section we study general bifurcation scenarios in the system of 
nonlinearly coupled oscillators (\ref{eq:gcpl1}), to be illustrated by
particular examples in the next section.

\subsection{Bifurcations in the Kuramoto--Sakaguchi model}
We start with the simplest case of \textit{linearly} coupled oscillators. Here
model (\ref{eq:gcpl1}) reduces to the standard Kuramoto-Sakaguchi model which we write as 
\begin{equation}
N\dot\theta_i=g_i(\theta_1,\ldots,\theta_N,\alpha)=-\sum_{j=1}^N\sin(\theta_i-\theta_j-\alpha) 
\label{eq:ks}
\end{equation}

\paragraph*{Equilibria.}
To describe the steady states of the corresponding system in
differences $\varphi_i=\theta_1-\theta_{1+i}$, we need to solves
the system of $N-1$ algebraic equations
\begin{equation}\label{AlgSys}
g_1(\theta_1,\dots,\theta_N,\alpha)-g_i(\theta_1,\dots,\theta_N,\alpha)=0,\quad
i=2,\cdots,N,
\end{equation}
where $\alpha$ is a scalar parameter. The next lemma helps
us to characterize the steady states of the system (\ref{eq:ks}).

\begin{lemma}\label{AlgSolutions}For
any $\alpha\in \T^1$, the set
$(\theta_1,\cdots,\theta_N)$ satisfies system of equations (\ref{AlgSys}) 
if and only if one of the following three conditions is fulfilled:\\
1) $\theta_1=\cdots=\theta_N$,\\
2) $\sum_{j=1}^N e^{i\theta_j}=0$,\\
3) $ \theta_1=\theta_2=\cdots =\theta_p\not =
\theta_{p+1}=\theta_{p+2}=\cdots =\theta_N,\quad p=1,\cdots,N-1$,
(plus all possible permutations).
\end{lemma}

This Lemma means that the only possible steady states are that of complete synchrony (one cluster), complete asynchrony, and of two clusters. 

\proof It is easy to check that states 1) -- 3) satisfy the system (\ref{AlgSys}). We will show that the
roots of the system (\ref{AlgSys}) satisfy 1) -- 3). We can re-write
(\ref{AlgSys}) in the following way:
\begin{equation}\label{algebraic1}
(\sin(\theta_1-\alpha)-\sin(\theta_i-\alpha))\sum_{j=1}^N
\cos\theta_j-
(\cos(\theta_1-\alpha)-\cos(\theta_i-\alpha))\sum_{j=1}^N
\sin\theta_j=0,
\end{equation}
where $i=2,\dots,N$. We consider four possible cases.\\
A. If $\sum_{j=1}^N \sin\theta_j=0$ and $\sum_{j=1}^N
\cos\theta_j=0$ simultaneously, then the condition 2) of the lemma
satisfies.\\
B. The next possible case is that of $\sum_{j=1}^N \sin\theta_j=0$ but
$\sum_{j=1}^N \cos\theta_j\not=0$. In this case (\ref{algebraic1})
implies
$$
\sin(\theta_1-\alpha)=\sin(\theta_i-\alpha), \quad i=2,\dots,N.
$$
The last system shows that we can obtain only two--cluster
solutions:
$$
\theta_i= \left\{
\begin{array}{ll}
\theta_1, & \ i=2,\dots,p,
\\[2mm]
-\theta_1+2\alpha+\pi, & \ i=p+1, \dots, N,
\end{array}
\right.
$$
which must satisfy equations
$$
p\sin\theta_1-(N-p)\sin(\theta_1-2\alpha)=0, \quad p=1,\dots,N.
$$
The last equations arise from $\sum_{j=1}^N\sin\theta_j=0$ and
they show that two--cluster states are possible only for some
values of
parameter $\alpha$ in this case. Note that the case $p=N$ corresponds to a 
one--cluster solution (condition 3 reduces to condition 1). \\
C. Consider the case, where $\sum_{j=1}^N \cos\theta_j=0$, and
$\sum_{j=1}^N \sin\theta_j\not=0$. As in the previous case we obtain
the possibility of two--cluster (or one--cluster, if $p=N$) states only:
$$
\theta_i= \left\{
\begin{array}{ll}
\theta_1, & \ i=2,\dots,p,
\\[2mm]
-\theta_1+2\alpha, & \ i=p+1, \dots, N,
\end{array}
\right.
$$
which satisfy conditions
$$
p\cos\theta_1+(N-p)\cos(\theta_1-2\alpha)=0,\quad p=1,\dots, N.
$$
D. Consider $(\theta_1,\cdots,\theta_n)$ such that $\sum_{j=1}^N
\sin\theta_j\not =0$ and $\sum_{j=1}^N \cos\theta_j\not =0$.
Denote $S:=\sum_{j=1}^N \sin\theta_j$, $C:=\sum_{j=1}^N
\cos\theta_j$, $s_j^{\alpha}:=\sin(\theta_j-\alpha)$,
$c_j^{\alpha}:=\cos(\theta_j-\alpha)$. Then equation
(\ref{algebraic1}) has the following form:
\begin{equation}\label{algebraic2}
(s_1^{\alpha}-s_i^{\alpha})C-(c_1^{\alpha}-c_i^{\alpha})S=0, \quad
i=2,\cdots,N.
\end{equation}
D1. Suppose that $s_1^{\alpha}-s_i^{\alpha}=0$ for all
$i=2,\cdots,N$. Then using inequality $S\not= 0$ we obtain
$c_1^{\alpha}-c_i^{\alpha}=0$ for $i=2,\cdots,N$. Equalities for
$s_j^{\alpha}$ and $c_j^{\alpha}$ considered together yield
$e^{i(\theta_j-\alpha)}-e^{i(\theta_1-\alpha)}=0$, $j=2,\dots,N$,
what means that all the values of $\theta_j$, $j=1,\dots,N$, are equal.\\
D2. Now let us consider another case, when there exists a number
$i_0$ such that $s_1^{\alpha}-s_{i_0}^{\alpha}\not=0$. Without
loss of generality we can set $i_0=2$. Then from the first of
equations (\ref{algebraic2}) we obtain
$$
C=(c_1^{\alpha}-c_2^{\alpha})S/(s_1^{\alpha}-s_2^{\alpha}).
$$
Substituting $C$ into the second equation of (\ref{algebraic2}),
we get
$$
S(s_1^{\alpha}-s_3^{\alpha})(c_1^{\alpha}-c_2^{\alpha})/(s_1^{\alpha}-s_2^{\alpha})-S(c_1^{\alpha}-c_3^{\alpha})=0.
$$
Using conditions $S\not =0$ and $(s_1^{\alpha}-s_2^{\alpha})\not
=0$, we obtain
$$
(s_1^{\alpha}-s_3^{\alpha})(c_1^{\alpha}-c_2^{\alpha})-(s_1^{\alpha}-s_2^{\alpha})(c_1^{\alpha}-c_3^{\alpha})=0,
$$
and then
$$
(s_1^{\alpha}c_3^{\alpha}-c_1^{\alpha}s_3^{\alpha})+
(s_2^{\alpha}c_1^{\alpha}-c_2^{\alpha}s_1^{\alpha})+
(s_3^{\alpha}c_2^{\alpha}-c_3^{\alpha}s_2^{\alpha})=0.
$$
After returning to the old notations and some transformations, we
obtain the expression
$$
\sin(\theta_1-\theta_3)+\sin(\theta_2-\theta_1)+\sin(\theta_3-\theta_2)=0,
$$
which already does not contain parameter $\alpha$. We provide the
last part of the proof  by contradiction. The case D supposes
that condition 2) is not valid. Now suppose that the
conditions 1) and 3) are not satisfied as well. This means that there exists
a solution $(\theta_1, \dots, \theta_N)$ of the system
(\ref{algebraic1}) such that at least three variables
$\theta_{i_1}$, $\theta_{i_2}$, $\theta_{i_3}$ of this solution
are not equal to each other. Without loss of generality we can set
$i_1=1$, $i_2=2$, $i_3=3$ because we can replace variables using
permutation (network has $S_N$ symmetry). Inequalities
$\theta_1\not =\theta_2$, $\theta_1\not =\theta_3$, $\theta_2\not
=\theta_3$ imply that
$$
\sin(\theta_1-\theta_3)+\sin(\theta_3-\theta_2)+\sin(\theta_2-\theta_1)=
$$
$$
=-4\sin\left(\frac{\theta_1-\theta_3}{2}\right)
\sin\left(\frac{\theta_3-\theta_2}{2}\right)
\sin\left(\frac{\theta_2-\theta_1}{2}\right) \not =0
$$
This contradiction proves validity of either 1) or 3).\\
D3. Consider a situation, when
$c_1^{\alpha}-c_{i_0}^{\alpha}\not=0$ for some number $i_0$. In
the same way as in the previous case D2 we prove that solutions of
(\ref{algebraic1}) satisfy one of the conditions 1) or 3).\\
Lemma is proved.

\paragraph*{Corollaries of Lemma 1.}
Lemma 1 implies that all steady states of the Kuramoto--Sakaguchi
system, in terms of the phase differences, are one-cluster, two-cluster, or completely desynchronized states. As the two-cluster states constitute straight lines (plus those obtained by permutations of the variables)
\begin{equation}\label{OnInvariantLine}
\varphi_1=\varphi_2=\cdots =\varphi_{p}\not
=\varphi_{p+1}=\varphi_{p+2}=\dots=\varphi_{N-1}=0, \quad
p=1,\dots,N-1,
\end{equation}
all bifurcations of
cluster steady states in this case are {\it one--dimensional} 
(in the sense that the normal forms are one-dimensional).
Furthermore, to study the existence of nontrivial cluster steady states  we only need to solve scalar algebraic
equations
\begin{equation}\label{EquationOn1DLines}
p\sin(\varphi_k-\alpha)+(N-p)\sin(\varphi_k+\alpha)-(N-2
p)\sin\alpha=0,\quad k=1,\dots, p.
\end{equation}
This equation has only two solutions on $T^1$: $\varphi_k=0$ and
\begin{equation}\label{Solution}
\varphi_k= \left\{
\begin{array}{ll}
\arccos\left(-\frac{2p(N-p)+(N^2-2p(N-p))\cos(2\alpha)}{N^2+2
p(N-p)(\cos(2\alpha)-1)}\right)
&\alpha\in\left[0;\frac{\pi}{2}\right)\cup
\left[\pi;\frac{3\pi}{2}\right),
\\[5mm]
-\arccos\left(-\frac{2p(N-p)+(N^2-2p(N-p))\cos(2\alpha)}{N^2+2
p(N-p)(\cos(2\alpha)-1)}\right)
&\alpha\in\left[\frac{\pi}{2};\pi\right)\cup
\left[\frac{3\pi}{2};2\pi\right).
\end{array}
\right.
\end{equation}
We can see that a bifurcation in the system (\ref{eq:ks}) occurs 
only when $\alpha=\pi/2$ and it
is transcritical. The bifurcation value of parameter $\alpha$
doesn't depend on the number of oscillators $N$ or on the cluster
partition (number $p$).

Note that in the case of a symmetric partition $N=2 p$, equations
(\ref{EquationOn1DLines}) have a very simple form
$$
2 p  \sin\varphi_k\cos\alpha=0.
$$
For this partition the only steady states  are $\varphi_k=0$ or $\varphi_k=\pi$, provided
$\alpha\not =\pm\pi/2$. There is no any bifurcation on
these lines for these values of the parameter. Vice verse, for  $\alpha=\pm\pi/2$ the
whole two-cluster invariant line in the case of symmetric partition consists of fixed points. These fixed points are degenerate saddles
(in the direction of lines with symmetry mentioned) and together
with their one--dimensional manifolds they build a set of
heteroclinic cycles.

As it follows from lemma \ref{AlgSolutions} and formula
(\ref{Solution}), the standard 
Kuramoto model of identical oscillators ($\alpha=0$) and the system
with coupling  $\alpha=\pm\pi/2$ have a simple structure of the steady states.
The standard Kuramoto model has only equilibria of two
types: (i) equilibria that compose the manifold $\mathcal{M}$ (with
vanishing order parameter) and
(ii) equilibria that have coordinates differences
$\theta_j-\theta_k$ equal to $0$ or to $\pm\pi$.
In the case $\alpha=\pm\pi/2$ all equilibria satisfy either (i) lie on the manifold  $\mathcal{M}$ or (ii) correspond to a completely synchronous
state, where $\theta_j=\theta_k$, $j,k=\overline{1,N}$, and $r=1$.

\subsection{Bifurcations in model of nonlinearly coupled oscillators}


The model of our main interest (system (\ref{eq:gcpl1}) or, equivalently, (\ref{PhaseDifferences})) 
differs from the Kuramoto-Sakaguchi model only by the nontrivial phase shift $\alpha$.
Fortunately, using Lemma 1 we can localize steady states in a system of equations  even more general
than (\ref{eq:ks}), with
r.h.s. containing an arbitrary scalar function
$$
\alpha={\alpha}(\theta_1,\dots,\theta_N,\beta),
$$
where $\beta$ is some vector of parameters
$\beta=(\beta_1,\dots,\beta_m)$, $m\ge 1$. To do this  we need to
describe all solutions of the algebraic system
\begin{equation}\label{AlgebricGeneralize}
g_1(\theta_1,\dots,\theta_N,{\alpha}(\theta_1,\dots,\theta_N,\beta))
-g_i(\theta_1,\dots,\theta_N,{\alpha}(\theta_1,\dots,\theta_N,\beta))=0,\quad
i=2,\cdots,N.
\end{equation}

\begin{lemma}\label{Alg2Solutions}
$(\theta_1,\cdots,\theta_N)$ satisfy system
(\ref{AlgebricGeneralize}) for any smooth scalar function
${\alpha}(\theta_1,\dots,\theta_N,\beta)$ and vector of
parameters $\beta\in \mathbb{R}^m$
if and only if they satisfy one of the following conditions:\\
1) $\theta_1=\cdots=\theta_N$,\\
2) $\sum_{j=1}^N e^{i\theta_j}=0$,\\
3) $ \theta_1=\theta_2=\cdots =\theta_p\not =
\theta_{p+1}=\theta_{p+2}=\cdots =\theta_N,\quad p=1,\cdots,N-1$,
up to permutations.
\end{lemma}

\proof Let us assume that conditions of lemma \ref{Alg2Solutions}
are violated for some fixed value of variables
$(\theta_1,\cdots,\theta_N)=(\theta_1^0,\cdots,\theta_N^0)$ and
parameters $\beta=\beta^0=(\beta_1^0,\cdots,\beta_m^0)$. Then
lemma \ref{AlgSolutions} is not valid for system
(\ref{AlgSys}) for the  fixed parameter value
$\alpha={\alpha}\left(\theta_1^0,\cdots,\theta_N^0,\beta^0\right)$.
This contradiction proves lemma \ref{Alg2Solutions}.

Note that here we don't require from function ${\alpha}$ (and
thus from coupling function of the whole system) any type of
symmetry. Nevertheless, all equilibria bifurcations are
one--dimensional and they occur on the straight lines which are
invariant for the system and are described by
(\ref{EquationOn1DLines}). However, in the paper we will consider
coupling function $g$ with permutation symmetry $S_N$ and will
describe bifurcations of the system using this symmetry property.

\begin{figure}
\centering
(a)\includegraphics[width=0.52\columnwidth]{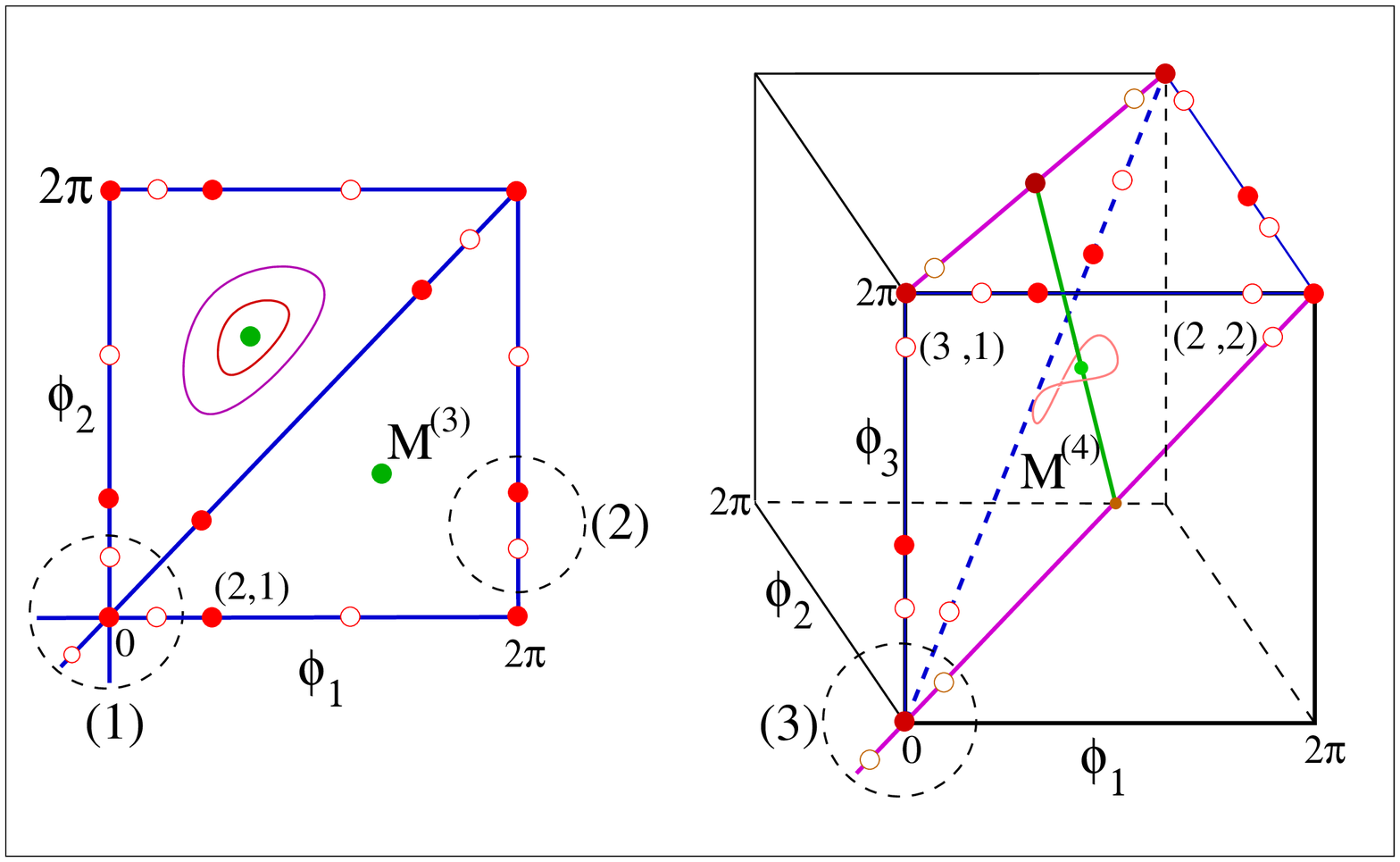}\hfill
(b)\includegraphics[width=0.38\columnwidth]{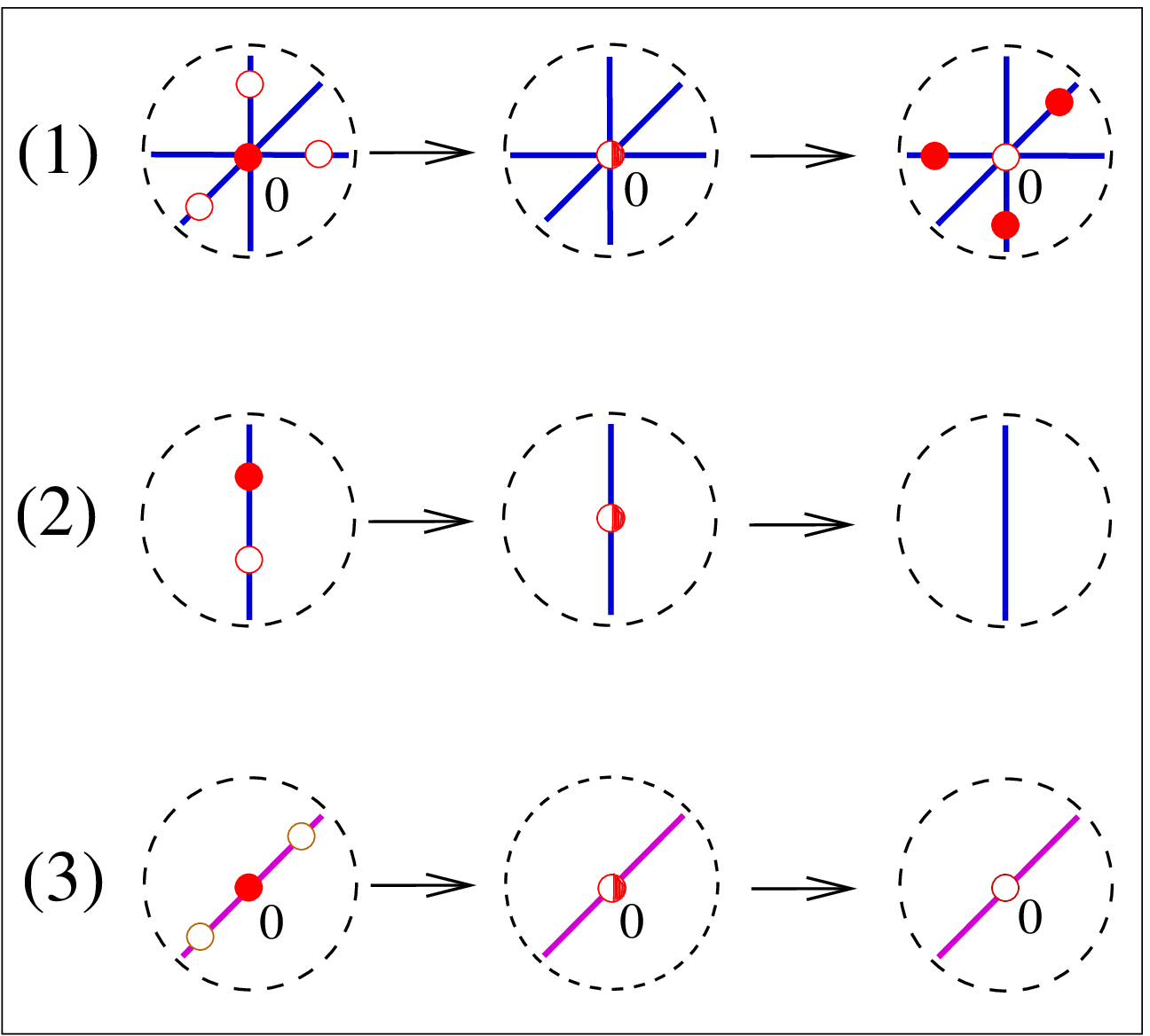}
\caption{Illustration of bifurcations of steady states for $N=3$ 
(left panel in (a), the system in terms of phase differences $\varphi$ is two-dimensional), 
and
$N=4$ (right panel in (a), the system in terms of phase differences $\varphi$ is 
three-dimensional). Panel (b) illustrates particular transitions in the selected 
regions of the phase space (see text for details).}
\label{fig:bifil}
\end{figure}

\begin{figure}
\centering
\includegraphics[width=0.7\columnwidth]{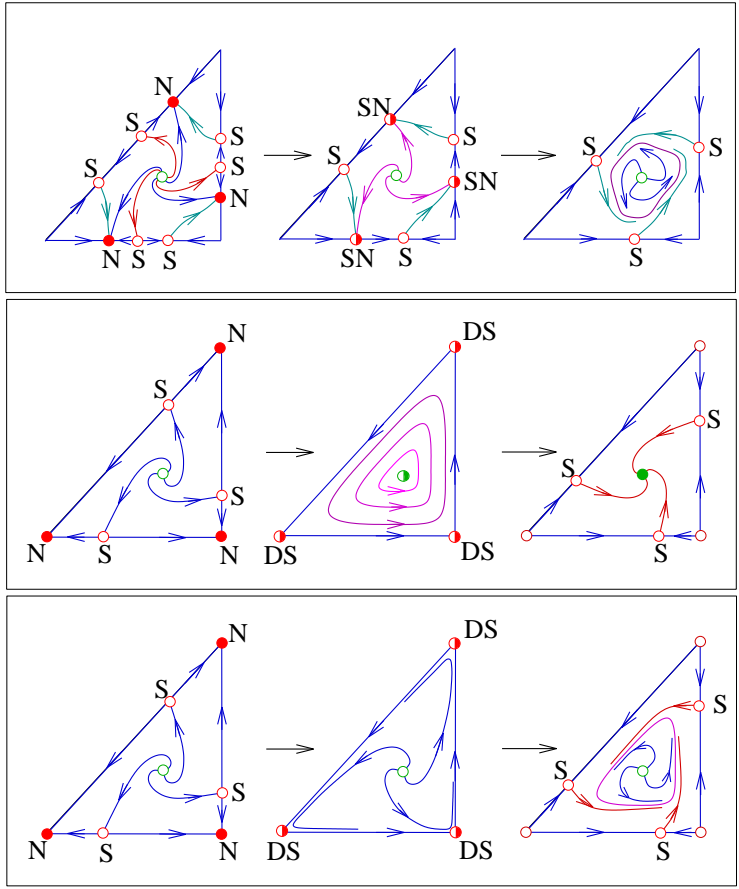}
\caption{Illustration of bifurcations via heteroclinic cycles. Upper panel: HC appears 
via a saddle-node bifurcation and gives rise to a limit cycle. Middle panel: The case of 
Kuramoto-Sakaguchi, here at the bifurcation point a family of neutral cycles exist, while
beyond it only a fully asynchronous steady state is stable. Bottom panel: HC appears 
via a transcritical bifurcation and gives rise to a limit cycle.}
\label{fig:hc}
\end{figure}

Lemma \ref{Alg2Solutions} shows that (like in the standard
Kuramoto-Sakaguchi model) all steady states of system
(\ref{PhaseDifferences}) (where
${\alpha}=\alpha(r,\beta)$) belong only to the invariant
manifold $\mathcal{M}$ or to clusters with isotropy $S_{p}\times
S_{N-p}$. In the latter case the problem reduces to 
solving scalar algebraic equations
\begin{equation}\label{EqOnInvLineGeneral}
p\sin(\varphi_k-\alpha(r(\varphi_k),\beta))+(N-p)\sin(\varphi_k+\alpha(r(\varphi_k),\beta))-(N-2
p)\sin(\alpha(r(\varphi_k),\beta))=0
\end{equation}
for these steady states. In these equations the mean field amplitude
$r$ is defined according to (\ref{OnInvariantLine}) and
it depends only on one variable $\varphi_k$, where $k=1,\dots,p$.
Also we can see that all steady state bifurcations have 1-dimensional normal forms
for the model. Below we describe these and other bifurcations, 
illustrating them with cases $N=3$ and $N=4$ (Figs.~\ref{fig:bifil},\ref{fig:hc}).

\paragraph*{Bifurcations of the completely synchronous state $\varphi_j=0$.}
The origin of the system (\ref{PhaseDifferences}) is
an equilibrium for any value of the function $\alpha(r,\beta)$.
Consider the Jacobian matrix of this system at the point
$\varphi_j=0$, $j=1,\cdots,N-1$. All eigenvalues of this matrix
have the same value:
$$
\lambda_i=-N\cos(\alpha(1,\beta)),\quad i=1,\cdots, N-1.
$$
This means that the origin of the system changes its stability
when $\alpha(1,\beta)=\pm\pi/2$. Also, as it was argued above, a
bifurcation must be one-dimensional on each of the invariant lines
with symmetry $S_p\times S_{N-p}$. This bifurcation can be either 
a transcritical or a pitchfork one. A pitchfork bifurcation can
happen only in the case of an even number of oscillators and this
bifurcation occurs along invariant lines with the symmetry
$S_{N/2}\times S_{N/2}$ as it was shown in the work of Ashwin and
Swift \cite{Ashwin92}.

Thus, a typical bifurcation of the completely synchronous state 
is a transcritical
bifurcation (see raw (1) in Fig.~\ref{fig:bifil}(b)). These bifurcations occur simultaneously on all
invariant lines with the isotropies $S_p\times S_{N-p}$, $p\not =
N/2$. Bifurcation parameters $\beta=(\beta_1,\dots,\beta_m)$ are
defined from the expression $\alpha(1,\beta)=\pm\pi/2$. The steady state 
at the bifurcation point is a degenerate saddle (all
eigenvalues of  the linearized system are zero).
$\sum_{j=1}^{[N+1]/2-1}C_N^j$ saddles, where [N] -- integer part
of $N$, meet together at the origin. The bifurcation changes
stability of the origin  along each of the one-dimensional directions.

A pitchfork bifurcation of the origin (see raw (3) in Fig.~\ref{fig:bifil}(b)) occurs simultaneously with the transcritical
bifurcation, when the number of oscillators is
even. Two saddles appear (disappear) from the origin (stable or
unstable) and move in opposite directions along the lines which
have $S_{N/2}\times S_{N/2}$ isotropy. In the case of an even $N$ 
these saddles are usually generators of
trajectories (one--dimensional manifolds) which can be parts of
heteroclinic cycles, under some additional conditions.

\paragraph*{Clusters and their bifurcations.}

To find all other steady states on the invariant lines with the
symmetries $S_p\times S_{N-p}$ we should solve appropriate
algebraic system (\ref{AlgebricGeneralize}) that satisfies
(\ref{OnInvariantLine}) -- this means that we need to solve one algebraic equation.
Typically, steady states appear (or disappear) by pairs on each
invariant lines for $\varphi_j\in(0, 2\pi)$, $j=1,\cdots,N-1$, and
this appearance (disappearance) corresponds to a saddle-node
bifurcation (see raw (2) in Fig.~\ref{fig:bifil}(b)). A saddle-node bifurcation that occurs in the
$(N-1)$-dimensional space (where our reduced system is considered)
leads to the appearance of two new points, i.e. of two new two-cluster states. 
These two points have opposite
stabilities along the one--dimensional manifold with isotropy
$S_p\times S_{N-p}$, but the same stabilities transversal to these one-dimensional manifolds.
In particular, one of these two newly appeared points
can be a stable or an unstable node. A stable node on the
one--dimensional invariant line corresponds to a two-cluster with
symmetry $S_p\times S_{N-p}$.

\paragraph*{Heteroclinic and limit cycles.}
Saddle steady states that appear in the transcritical and saddle-node
bifurcations described above may have unstable manifolds that are
connected to each other, thus constituting a heteroclinic cycle (see also a 
similar structure described in \cite{Ashwin-etal-08}). When the heteroclinic cycle disappears, a usual 
limit cycle may appear corresponding to a periodic non-synchronous regime in 
system (\ref{PhaseDifferences}). We illustrate two types of such a bifurcation in Fig.~\ref{fig:hc}. 
In the upper panel we show an appearance of a limit cycle via a heteroclinic one, 
that appears at the saddle-node collision. Generation of a limit
cycle by a saddle--node bifurcation via a heteroclinic cycle  is a typical
situation in the system (\ref{PhaseDifferences}). In
the case of an even number of oscillators, saddle--node bifurcations on
the invariant lines can give possibility to connect different
one--dimensional manifolds of saddles (pairs of saddles) generated
by pitchfork bifurcation from the origin. 
The bottom panel illustrates a heteroclinic cycle appearing at a transcritical bifurcation at the origin. Heteroclinic
(or homoclinic) cycle consists of the origin point and loops of
$S_p\times S_{N-p}$ invariant lines. The middle panel in Fig.~\ref{fig:hc} shows the same transcritical bifurcation in the Kuramoto-Sakaguchi model.

Another possibilities  of a limit cycle to appear 
are an Andronov--Hopf bifurcation of the point of $\mathcal{M}$ (we will
show this below), and a saddle--node bifurcation of two limit cycles.
The existence of more complicated structures such as of a
quasi--periodic torus or chaotic attractors is impossible for the
system in phase differences (\ref{PhaseDifferences}). This
follows from the Watanabe-Strogatz theory~\cite{Watanabe-Strogatz-94}. 
As it was shown in \cite{Watanabe-Strogatz-94,Pikovsky-Rosenblum-09}, the system
(\ref{eq:gcpl1}) can be reduced to a skew three--dimensional
system where  the equation for one variable fully depends on two
other ones. Thus, the dynamics of the two ``driving'' variables can be at most periodic, and the full dynamics at most quasiperiodic. In the terms of variables we use here, the ``driving'' variables correspond to phase differences $\varphi_k$, their dynamics thus can be at most periodic. The full dynamics of phases $\theta_k$ includes one more integration and can be at most quasiperiodic. 

\paragraph*{Multistability.}
If a saddle-node bifurcation generates a stable node while a stable node at the origin
still exists, we obtain a bistability of a fully synchronized
and two-cluster regimes. Note that depending on
function $\alpha(r,\beta)$, we can obtain many stable nodes on
the invariant lines, resulting in a multistability of synchronous and different two-cluster states.

One can also observe a coexistence of limit cycles appeared via
different saddle--node bifurcations accompanied by heteroclinic cycles. 
Part of these cycles are
stable but other ones are not. 

\paragraph*{Attractors.}
As a result, one can observe the following types of possible stable
regimes or their combinations in system (\ref{PhaseDifferences}):

{\it
1) Complete synchrony $\varphi_j=0$, $j=1,\cdots N$.

2) Two--cluster regime with symmetry $S_p\times S_{N-p}$.

3) Limit cycle.

4) Heteroclinic cycle.

5) Manifold $M^{(N)}$.
}

\paragraph*{Stability of $\mathcal{M}$.}
Consider the invariant set $\mathcal{M}$. This set is $(N-3)$--dimensional
in $\T^{N-1}$ and consists of steady states of the system. To
describe local bifurcations,
we need to consider the property of Jacobin matrix
$$
J=J(\varphi_1,\dots,\varphi_{N-1},\alpha(r,\beta))=\frac{\partial
(g_1(\varphi_1,\dots,\varphi_{N-1},\alpha),\dots,g_{N-1}(\varphi_1,\dots,\varphi_{N-1},\alpha)
)}{\partial (\varphi_1,\dots,\varphi_{N-1})}
$$
on the points of the manifold $\mathcal{M}$. We will show that $N-3$ eigenvalues
of Jacobian vanish, so there is no any motion
inside the manifold.

\begin{lemma}\label{RankJacobian}
Jacobian rank of the system (\ref{PhaseDifferences}) is:
$$
{\rm rank}(J)= \left\{
\begin{array}{ll}
1, & \mbox{for 2--clusters with symmetry} \ S_{N/2}\times
S_{N/2}, \\
2, & \mbox{in other cases}.
\end{array}
\right.
$$
\end{lemma}

\proof Jacobian matrix $J$ has the elements
$$
J_{kk}=\frac{\partial g_k}{\partial\varphi_k}=- \left[
\cos(\varphi_k-\alpha)
-\frac{\partial\alpha}{\partial\varphi_k}\sum_{j=1}^{N-1}\cos(\varphi_j-\alpha)+
\right.
$$
$$
+ \left. \left(1+ \frac{\partial\alpha}{\partial\varphi_k}\right)
\left(\cos(\varphi_k+\alpha)+ \sum_{j=1,j\not
=k}^{N-1}\cos(\varphi_k-\varphi_j+\alpha)\right) \right],
$$
$$
J_{ki}=\frac{\partial g_k}{\partial\varphi_i}=- \left[
\cos(\varphi_i-\alpha) -\cos(\varphi_k-\varphi_i+\alpha)
-\frac{\partial\alpha}{\partial\varphi_i}\sum_{j=1}^{N-1}\cos(\varphi_j-\alpha)+
\right.
$$
$$
+ \left. \frac{\partial\alpha}{\partial\varphi_i}
\left(\cos(\varphi_k+\alpha)+ \sum_{j=1,j\not
=k}^{N-1}\cos(\varphi_k-\varphi_j+\alpha)\right) \right],
$$
Since we consider manifold $\mathcal{M}$, then using (\ref{Manifold})
and (\ref{substitution}) we obtain
$$
\cos\alpha+\sum_{j=1}^{N-1}\cos(\varphi_j-\alpha)=0
$$
and
$$
\cos(\varphi_k+\alpha)+\cos\alpha+ \sum_{j=1,j\not
=k}^{N-1}\cos(\varphi_k-\varphi_j+\alpha)=0.
$$
Thus in this case elements of Jacobian matrix are
$$
\frac{\partial g_k}{\partial\varphi_j}\bigg|_{M^{(N)}}=
\cos(\varphi_j-\varphi_k-\alpha(0,\beta))-\cos(\varphi_j-\alpha(0,\beta)),
\quad j,k=1,\dots,N-1.
$$

Denote each column of matrix $J$ by $\overline{J}_k$,
$k=1,\dots,N-1$. To prove that rank of matrix is not greater than
two, we need to show that there exists the linear dependence
between any three columns $\overline{J}_i$, $\overline{J}_k$,
$\overline{J}_l$ of matrix $J$, i.e. there exist two scalar
functions $\gamma_j$ and $\gamma_k$ such that
$$
\gamma_j \overline{J}_j+\gamma_k \overline{J}_k=\overline{J}_l.
$$
One can check that the last expression is satisfied with functions
$$
\gamma_j=\frac{\sin(\varphi_l-\varphi_k)}{\sin(\varphi_j-\varphi_k)},\quad
\gamma_k=\frac{\sin(\varphi_l-\varphi_j)}{\sin(\varphi_k-\varphi_j)},
$$
when $\varphi_j\not =\varphi_k$. Thus ${\rm rank}(J)\le 2$.

We can rewrite equation for vectors in the form
$$
\sin(\varphi_k-\varphi_l)\overline{J}_j+\sin(\varphi_l-\varphi_j)\overline{J}_k+
\sin(\varphi_j-\varphi_k)\overline{J}_l=0.
$$
All coefficients are not equal to zero in this expression when
$\varphi_i\not=\varphi_k\not=\varphi_l$. Thus ${\rm rank}(J)$ is
not less than two when the system has at least a three--cluster
regime. Then ${\rm rank}(J)=2$ for three--or--more cluster
regimes.

In the case of even number of oscillators $N=2 p$ the system can
have two--cluster states with symmetry $S_{N/2}\times S_{N/2}$
that belong to invariant manifold ${\cal M}$. This means that in
the last equation one coefficient is equal to zero that implies
${\rm rank}(J)=1$.
The lemma is proved.

The lemma shows that the Jacobian has $N-3$ eigenvalues equal to zero
at the points of manifold $\mathcal{M}$. However, as it was shown, the rank
of the Jacobian depends on values of the variables 
(i.e. on the coordinates of points
on the manifold). Thus, to find the eigenvalues of $J$ we need to
consider not only $2\times 2$ minor of Jacobian matrix but the
whole matrix. Each of the eigenvalues is a function of $N-3$ variables
in the points of the manifold. Let us express the last two
variables $\varphi_{N-2}$ and $\varphi_{N-1}$ as the functions of
the variables $\varphi_1,\dots,\varphi_{N-3}$ using expressions of
real and imaginary parts of (\ref{Manifold}). Then we obtain:
$$
\varphi_{N-2}=\arctan\left(\frac{f_2}{f_1}\right)-\frac{1}{2}\arccos\left(
\frac{f_1^{2}+f_2^{2}}{2}-1 \right)+\frac{\pi}{2} (1-{\rm sign}
(f_1)),
$$
$$
\varphi_{N-1}=\arctan\left(\frac{f_2}{f_1}\right)+\frac{1}{2}\arccos\left(
\frac{f_1^{2}+f_2^{2}}{2}-1 \right)+\frac{\pi}{2} (1-{\rm sign}
(f_1)),
$$
where
$$
f_1(\varphi_1,\dots,\varphi_{N-3})=-1-\sum_{j=1}^{N-3}\cos\varphi_j,\quad
f_2(\varphi_1,\dots,\varphi_{N-3})=-\sum_{j=1}^{N-3}\sin\varphi_j.
$$

In the case of a uniform distribution of oscillators on the circle 
(splay state according to terminology used for Kuramoto model~\cite{Watanabe-Strogatz-94}),
that is when
$$
\varphi_1=\frac{2\pi}{N},\quad \varphi_j=j\varphi_1, \quad
j=2,\dots, N-1,
$$
the eigenvalues of the Jacobian are
$$
\lambda_{N-2,N-1}=\frac{N}{2}(\cos\alpha\pm i \sin\alpha).
$$
In general case the eigenvalues are
$$
\lambda_{N-2,N-1}(\varphi_1,\dots,\varphi_{N-3})=\frac{N}{2}\left(\cos\alpha\pm
\sqrt{\cos^2\alpha-h^2(\varphi_1,\dots,\varphi_{N-3})}\right),
$$
where $|h(\varphi_1,\dots,\varphi_{N-3})|\le1$ is some enough
complicate smooth function. Therefore, we obtain an Andronov--Hopf
bifurcation when $\alpha=\pm\pi/2$. This bifurcation happens
simultaneously in each point of manifold $\mathcal{M}$ except for the
points with isotropy $S_{N/2}\times S_{N/2}$ where function
$h(\varphi_1,\dots,\varphi_{N-3})=0$.

In the case of three coupled phase oscillators, the zero--dimensional
manifold $\mathcal{M}^{(3)}$ consists of two points $(2\pi/3,4\pi/3)$ and
$(4\pi/3,2\pi/3)$. At the point of an Andronov--Hopf bifurcation each
of these two points changes its stability and generates supercritically (or
destroys subcritically) a
limit cycle. With a further variation of a parameter, 
this limit cycle can grow in amplitude and disappear, either in
a saddle-node/heteroclinic bifurcation, or via  a
saddle-node bifurcation of two limit cycles.

More nontrivial situations can happen in the case of four globally
coupled oscillators. Invariant manifold $\mathcal{M}^{(4)}$ in this case
consists of six straight lines.
Coordinates of such lines are
$(\varphi_j,\pi,\varphi_j+\pi)$ up to permutations. Invariant
manifold has $Z_2$ isotropy. The function $h(\varphi_j)=\sin\varphi_j$
appears
in the expressions for the eigenvalues of the Jacobian matrix.
An Andronov--Hopf bifurcation happens simultaneously in each of
the manifold points. Thus, we obtain a two dimensional surface that
consists of limit cycles. Noteworthy, the bifurcation differs
from a Neimark--Sacker bifurcation. Each of this cycles is
attractive (repulsive) only inside the surfaces described by
Watanabe--Strogatz theory \cite{Watanabe-Strogatz-94}, in other direction it is
neutral. Possible way of this two-dimensional surface to disappear a is
saddle-node heteroclinic bifurcation on invariant lines described.
A two--dimension set of heteroclinic cycles occurs at the point of
bifurcation (such a set was shown in figure 10 in \cite{Ashwin-etal-08}).

Another possibility is disappearance (appearance) of two limit
cycles in a saddle--node bifurcation of cycles. We can note that
such a bifurcation happens for each pair of limit cycles which
belong to different two--dimensional sets of cycles. Such
a bifurcation happens also inside the Watanabe--Strogatz surfaces. Thus
we obtain a saddle--node bifurcation of two--dimensional surfaces,
one of them is stable and other is unstable.

\section{Nonlinearly coupled oscillators with quadratic phase nonlinearity}
\label{sec:ncowqpn}

As an example of application of general picture outlined above we 
consider the model (\ref{eq:gcpl1}) with particular dependence of the
phase shift on the amplitude of the mean field~\cite{Pikovsky-Rosenblum-09}:
\begin{equation}
\alpha=\alpha(r,\beta)=\beta_1+\beta_2 r^2.
\label{eq:example}
\end{equation}
Here the two--dimensional space of parameters $(\beta_1,\beta_2)$ 
is a cylinder $\mathbb{R}\times
\mathbb{T}\supset (\beta_2,\beta_1)$, because the r.h.s.  of
the equations are $2\pi$--periodic with respect to $\beta_1$. The oddness of the r.h.s. of the system implies
the symmetry of the parameter plane $(\beta_1,\beta_2)\to(-\beta_1,-\beta_2)$.

According to the consideration above, two types of bifurcation happen when
$$
\cos(\alpha)=\cos(\beta_1+\beta_2 r^2)=0.
$$

One possible bifurcation is an Andronov--Hopf bifurcation (AH) on
the invariant manifold $\mathcal{M}$. Since $r=0$ on the manifold,
then we obtain two straight bifurcation lines
\begin{equation}
\beta_1=\pi/2\quad \mbox{and}\quad \beta_1=3\pi/2
\label{eq:bifl1}
\end{equation}
on the parameter cylinder. The line $\beta_1=\pi/2$ corresponds to
a supercritical Andronov--Hopf bifurcation and the line
$\beta_1=3\pi/2$ corresponds to a subcritical one. Manifold
$\mathcal{M}$ is stable if $\beta_1\in (\pi/2, \ 3\pi/2)$ and it is
unstable if $\beta_1\in [0,\pi/2)\cup (3\pi/2,2\pi)$.

Another possible bifurcation is a transcritical bifurcation (TC) at the origin
($\varphi_j=0$, $j=1,\dots,N-1$) along each of the invariant lines
with the symmetry $S_p\times S_{N-p}$, $p=1,\dots,N-1$. A
pitchfork bifurcation (PF) at the origin occurs, simultaneously with
the transcritical one,  along the invariant lines with the
symmetry $S_{N/2}\times S_{N/2}$ for the system with even number
of oscillators. At this bifurcation point the origin 
is a degenerate saddle with $N-1$ zero eigenvalues. The origin
point of the system corresponds to full synchronization state
where order parameter $r=1$. Therefore, straight lines
\begin{equation}
\beta_1+\beta_2=\pi/2+\pi m,\quad m\in\mathbb{Z},
\label{eq:bifl2}
\end{equation}
correspond to the transcritical (TC) or to the transcritical--pitchfork (TC/PF) bifurcations in the
parametric space. The origin (i.e. the regime of full synchrony)
is stable when $\beta_1+\beta_2\in
(-\pi/2+2\pi m, \ \pi/2+2\pi m)$, $m\in\mathbb{Z}$, and it is
unstable when $\beta_1+\beta_2\in (\pi/2+2\pi m, \ 3\pi/2+2\pi
m)$.

These two types of bifurcation are independent of the number of oscillators, so 
the grids of straight bifurcation lines (\ref{eq:bifl1},\ref{eq:bifl2}) are present at any
bifurcation diagram model (\ref{eq:example}) (Figs.~\ref{fig:3Diagr},\;\ref{fig:45Diagr}). Other
bifurcations of the fixed points occur only on invariant lines
(\ref{OnInvariantLine}) and they all are of the saddle--node type.
The expressions for the order parameter on the invariant lines
(\ref{OnInvariantLine}) are
$$
r^2=r^2(p,N-p)=\frac{1}{N^2} (2
p(N-p)\cos(\varphi_k)+(N-1)^2-2(p-1)(N-p-1)+1),
$$
where $\varphi_k$, $k=1, \dots,p$, is a variable that changes along
invariant lines with $S_p\times S_{N-p}$ isotropy. To find the
coordinates of the steady states we need to solve equation
(\ref{EqOnInvLineGeneral}) with this expression for the order parameter. 
This expression simplyfies in the case of even number oscillators to 
$$
r^2(p,p)=\frac{1}{2}(\cos(\varphi_k)+1),\quad k=1,\dots,p\;,
$$
and is independent of the number of oscillators. Thus, it describes
appearance (disappearance) of two points on the invariant line with symmetry
$S_{N/2}\times S_{N/2}$  after each pitchfork bifurcation at
the origin. The coordinates of these points (which are saddles) on
the invariant lines are then
$$
\varphi_k=\pm\arccos\left( \frac{1}{\beta_2} (\pi(1+2
m)-2\beta_1-\beta_2) \right),\quad m\in\mathbb{Z}.
$$
These symmetric points are important because they are the basis
for heteroclinic cycles in cases of even number of oscillators.

For any number of oscillators, together with  the grid of the straight lines
describing Andronov--Hopf and transcritical bifurcations (\ref{eq:bifl1},\ref{eq:bifl2}),
there are a lot of bifurcation lines that correspond to
saddle--node bifurcations (SN) on invariant lines. Some of these lines
correspond to heteroclinic bifurcations (HC or HC(SN)), provided  some 
additional
conditions are satisfied. At these saddle--node/heteroclinic bifurcations 
 limit cycles appear. Let us fix some parameter value $\beta_1\not=\pm\pi/2$
and increase parameter $\beta_2$. Then new and new saddle--node
bifurcations occur on the invariant lines, steady states
appear at these bifurcations in  such a way that their stability
alternate along this line. Stability of appearing heteroclinic cycles
also alternates with increasing of the parameter $\beta_2$.
Thus, stabilities of limit cycles that move inside each invariant
region after such a bifurcation also alternate. The period and the amplitude
of each of the limit cycles decrease when parameter $\beta_2$
increases, but no bifurcation of small limit cycle can happen
because we demanded that  $\beta_1$ was not equal to $\pm\pi/2$
where such a bifurcation only could occur.

The number of the hyperbolic steady states increases with increasing of
parameter $\beta_2$. These steady states tend to concentrate near the center part of the
invariant lines. A saddle--node/heteroclinic bifurcation on the line
with symmetry $S_1\times S_{N-1}$ usually occurs close to this 
central region, where the coordinate $\varphi$ of the saddle--node
point is close to $\pi$. Thus,  we can approximately calculate that such
bifurcation occurs at
$$
\beta_2 \approx \frac{N^2}{(N-2)^2}(\pi m-\beta_1), \quad
m\in\mathbb{Z}.
$$
The lines of a saddle--node/heteroclinic bifurcation alternates on the
bifurcation cylinder. Each heteroclinic cycle generates a limit
cycle (or a set of limit cycles for 4 and more oscillators) with the
same stability. The period and the size of each cycle decrease with
increasing of parameter $\beta_2$. Stable and unstable limit cycles
enwrap each other inside invariant region, and alternate.
A saddle--node bifurcation of limit cycles is impossible for the
system considered because of monotonic increase of cycles
sizes (it would be, however, possible for more complex than (\ref{eq:example}) 
 dependencies $\alpha(r)$, e.g. for
$\alpha=\beta_1+\beta_2r^2+\beta_3r^4$).

Noteworthy, the first saddle--node bifurcation in the system usually
happens when $\beta_2<\pi/2-\beta_1$, and this bifurcation can
generate a stable node. Then we obtain multistability of the fully
synchronous state (the origin where $r=1$) and the
stable two--cluster states. If
the first transcritical bifurcation (when $\beta_2>\pi/2-\beta_1$)
doesn't produce a stable heteroclinic cycles, 
then the two--cluster states are the only attractors in
the system. The stable nodes accumulate on the invariant line with
increasing $\beta_2$, thus we obtain a multistability of
two--cluster states when $\beta_2$ is large enough. The appearance
of stable limit cycles after saddle--node/heteroclinic bifurcation
eliminates one stable node (on each invariant line with the same
symmetry). However, since a heteroclinic bifurcation happens more rarely
than simple saddle--node bifurcations of two points, then the coexistence of a stable
limit cycle with two--cluster states is typical for the system.
Therefore, we can obtain multistability of all possible attractors
in the system: full synchronous state, two--cluster state (with
different order parameters), limit cycle of phase differences,
heteroclinic cycle, and invariant manifold $\mathcal{M}$.

\subsection{Three interacting oscillators}
\begin{figure}
\centering
\includegraphics[width=0.5\columnwidth]{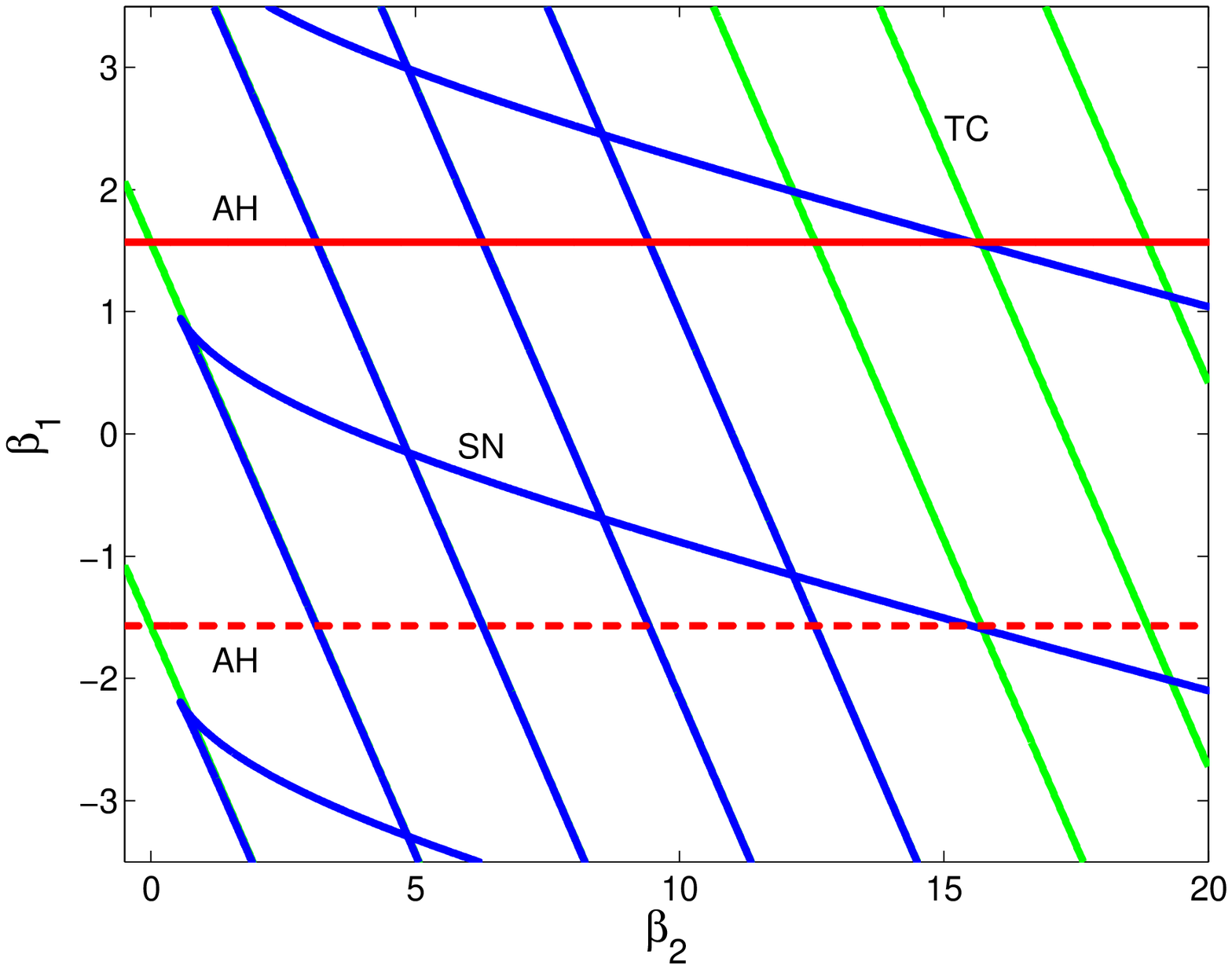}
\includegraphics[width=0.43\columnwidth]{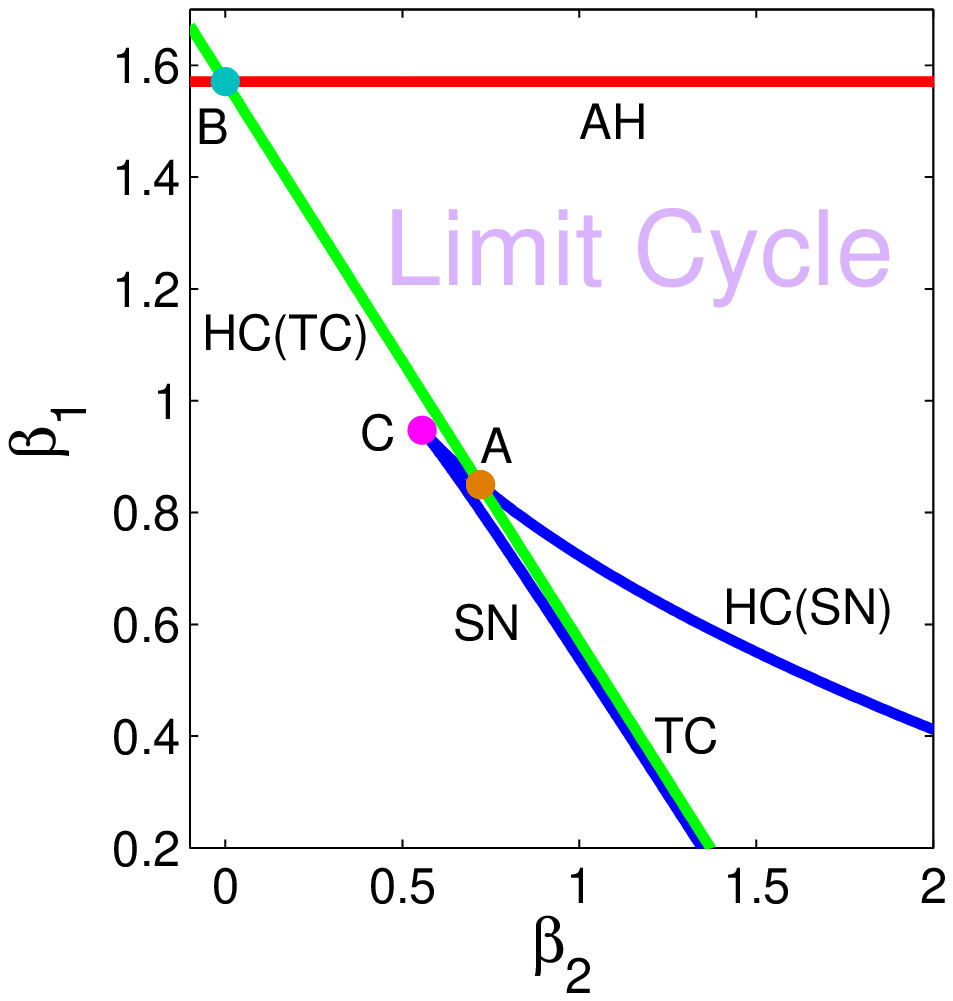}
\caption{Bifurcation diagram for $N=3$ oscillators
in the $(\beta_2,\beta_1)$ parametric plane (cf. Fig.~\ref{fig:bifil}(a)). Right panel is an enlargement 
of the central part of the left one. TC
--- transcritical bifurcation (see Fig.~\ref{fig:bifil}(b)(1)), SN --- saddle--node bifurcation
(see Fig.~\ref{fig:bifil}(b)(2)),
AH --- supercritical Andronov--Hopf bifurcation (see Fig.~\ref{fig:bifil}(b)(4)), HC(SN), HC(TC)
heteroclinic bifurcations (see Fig.~\ref{fig:hc}). Points  A, B and C are codimension--two bifurcation
points. The region where a stable limit cycle exists (right panel)
is surrounded by a supercritical AH
bifurcation line and two lines of heteroclinic bifurcations of different types.}
\label{fig:3Diagr}
\end{figure}

\begin{figure}
\begin{center}
\includegraphics[width=0.7\columnwidth]{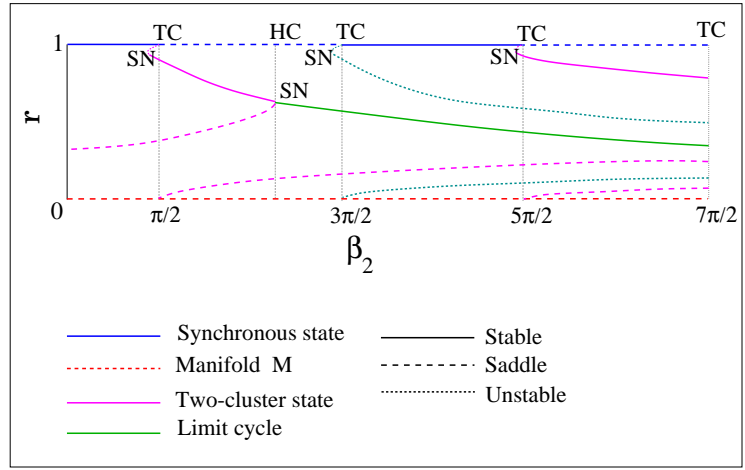}
\end{center}
\caption{Schematic bifurcation diagram for
the $(\beta_2,r)$ parametric plane and the case $\alpha=\beta_2 r^2$. TC ---
transcritical bifurcation, HC
--- heteroclinic cycle (bifurcation), SN --- saddle--node bifurcation. }\label{DiagrRadius}
\end{figure}

For three oscillators interacting according to phase shift (\ref{eq:example}), the bifurcation
diagram is depicted in Fig.~\ref{fig:3Diagr}. The corresponding bifurcations have been already illustrated in 
Figs.~\ref{fig:bifil},\ref{fig:hc} above. According to this bifurcation diagram, we show in 
Fig.~\ref{DiagrRadius} a schematic dependence of the synchronization states in the system 
as parameter $\beta_2$ changes while $\beta_1=0$. One can see that the basic transition in terms of 
phase differences is:
Full synchrony $\to$ two-cluster state $\to$ periodic oscillations. 
The first transition is with hysteresis (i.e. in some small region
of parameters full synchrony and two-cluster state coexist), and the second transition is via 
heteroclinic connection. On the diagram Fig.~\ref{fig:3Diagr} several codimension-2 points are marked, 
we will discuss them in the next subsection.

\subsection{Four and more coupled oscillators}

\begin{figure}
\centering
\includegraphics[width=0.46\columnwidth]{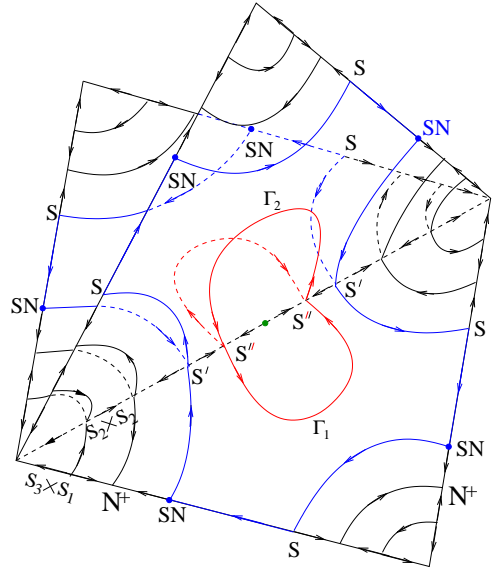}
\caption{Schematic phase portrait for $N=4$
coupled oscillators, showing only invariant lines corresponding to cluster states $1+3$ and $2+2$,
and planes connecting them. We illustrate 
a coexistence of two types of heteroclinic cycles. One
unstable HC (blue) is shown at the bifurcation point, another stable HC (red) 
is beyond its bifurcation.}
\label{fig:4cycles}
\end{figure}

\begin{figure}
\centering
\includegraphics[width=0.7\columnwidth]{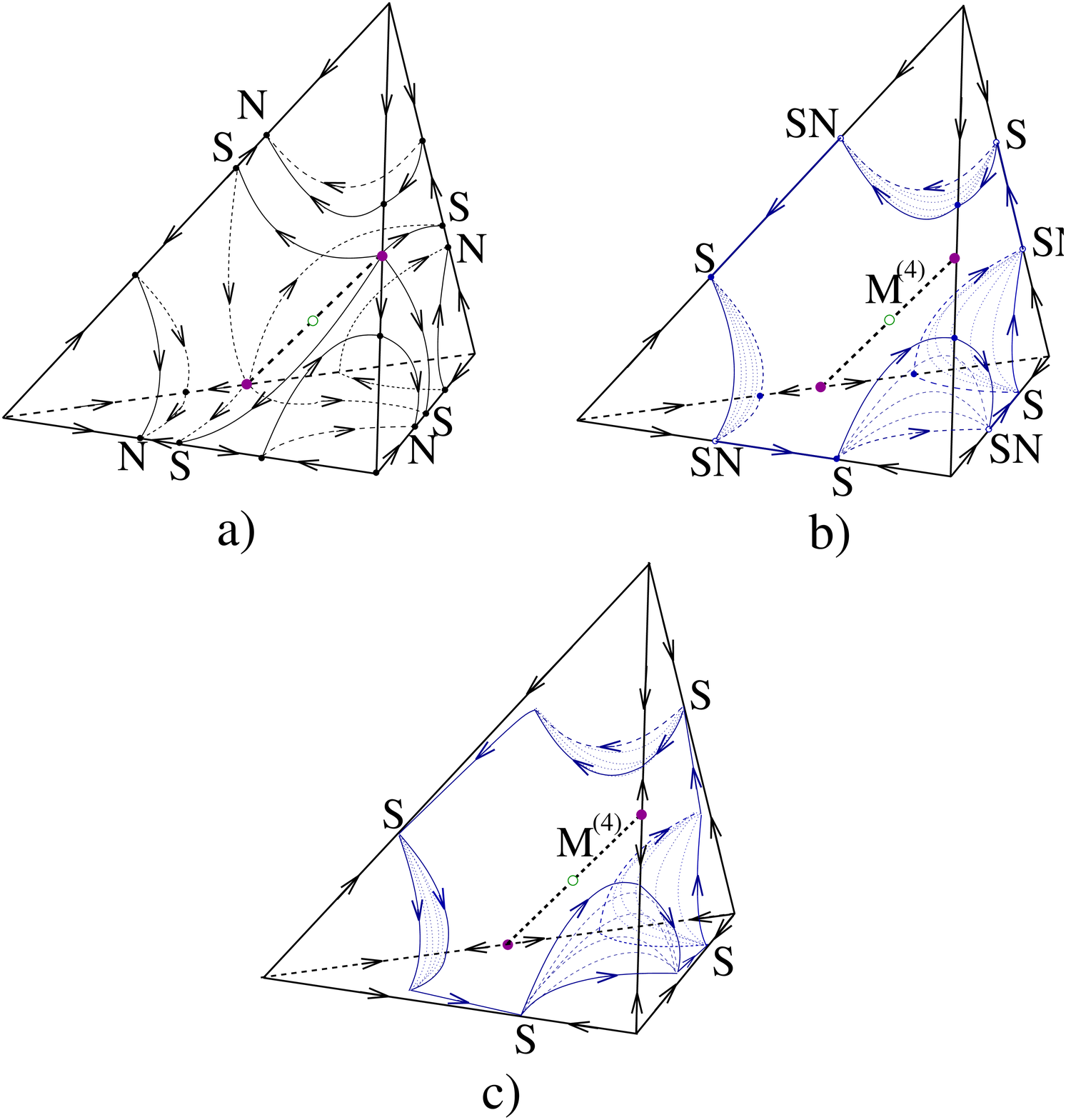}
\caption{Schematic diagrams of the
saddle--node/heteroclinic bifurcation for 4 globally coupled
oscillators. a) -- prior bifurcation, b) -- at the bifurcation point, c) beyond
bifurcation.}
\label{fig:4diags}
\end{figure}

In the Figures \ref{fig:4cycles} and \ref{fig:4diags} we show
schematically a saddle--node/heteroclinic
bifurcation for the case $N=4$.  An unstable heteroclinic cycle
(Figure \ref{fig:4cycles}) consists of ten fixed points and ten
1--dimensional invariant manifolds that connect these points. Four
saddle--node bifurcations happen simultaneously on four $S_1\times
S_3$ invariant lines. This 
heteroclinic cycle includes also  two
saddles $S'$ that belong to $S_2\times S_2$ invariant line. Thus,
the system of four oscillators,  that moves along this heteroclinic cycle,
shows temporary switches between $1+3$ and $2+2$
clustering. This unstable heteroclinic cycle is robust and
it will exist also beyond the saddle-node bifurcation; however  it will consist of two
lines connecting  $S'$ only, like the stable cycle depicted in the figure.
This stable heteroclinic cycle is shown inside
the unstable one. It consists of two saddles $S''$ and two connecting lines
$\Gamma_1$, $\Gamma_2$. The stable heteroclinic cycle appears at a
saddle--node bifurcation on the invariant line in the same way as the
unstable one, only for a smaller value of parameter $\beta_2$. The
next heteroclinic bifurcation will occur after merging of stable
node $N^+$ and saddle $S$ and it will produce causes stable heteroclinic cycle.

Figure \ref{fig:4diags} shows an appearance of a
2--dimensional sets of stable limit cycles inside one invariant
region. Four pairs of saddles $S$ and stable nodes $N$ (Fig.~\ref{fig:4diags} a) ) 
collide and create 2--dimensional sets
of heteroclinic cycles (Fig.~\ref{fig:4diags} b) ). Beyond the
bifurcation, when saddle--node points $SN$ disappear, two
heteroclinic cycles appear, with  2--dimensional sets of limit cycles
between them (Fig.~\ref{fig:4diags} c) ). The set of
limit cycles surrounds the 1--dimensional invariant set $\mathcal{M}$. This
set of limit cycles shrinks as parameter $\beta_2$ increases, but it never
reaches manifold $\mathcal{M}$.

The heteroclinic cycles presented in Fig.~\ref{fig:4cycles} lie on invariant surfaces and correspond 
to switches between the cluster states. They are borders of sets of limit cycles that exist \textit{inside} 
the bulk of the phase space (that is bounded by the 
invariant lines and surfaces), and enwrap the manifold $\mathcal{M}$. To an unstable HC 
corresponds a cylindrical set of unstable limit cycles, and to a stable HC corresponds
a cylindrical set of 
stable limit cycles. In this way the structure of heteroclinic cycles determines the overall structure of
the trajectories also outside of invariant manifolds.

\begin{figure}
\centering
(a)\includegraphics[width=0.46\columnwidth]{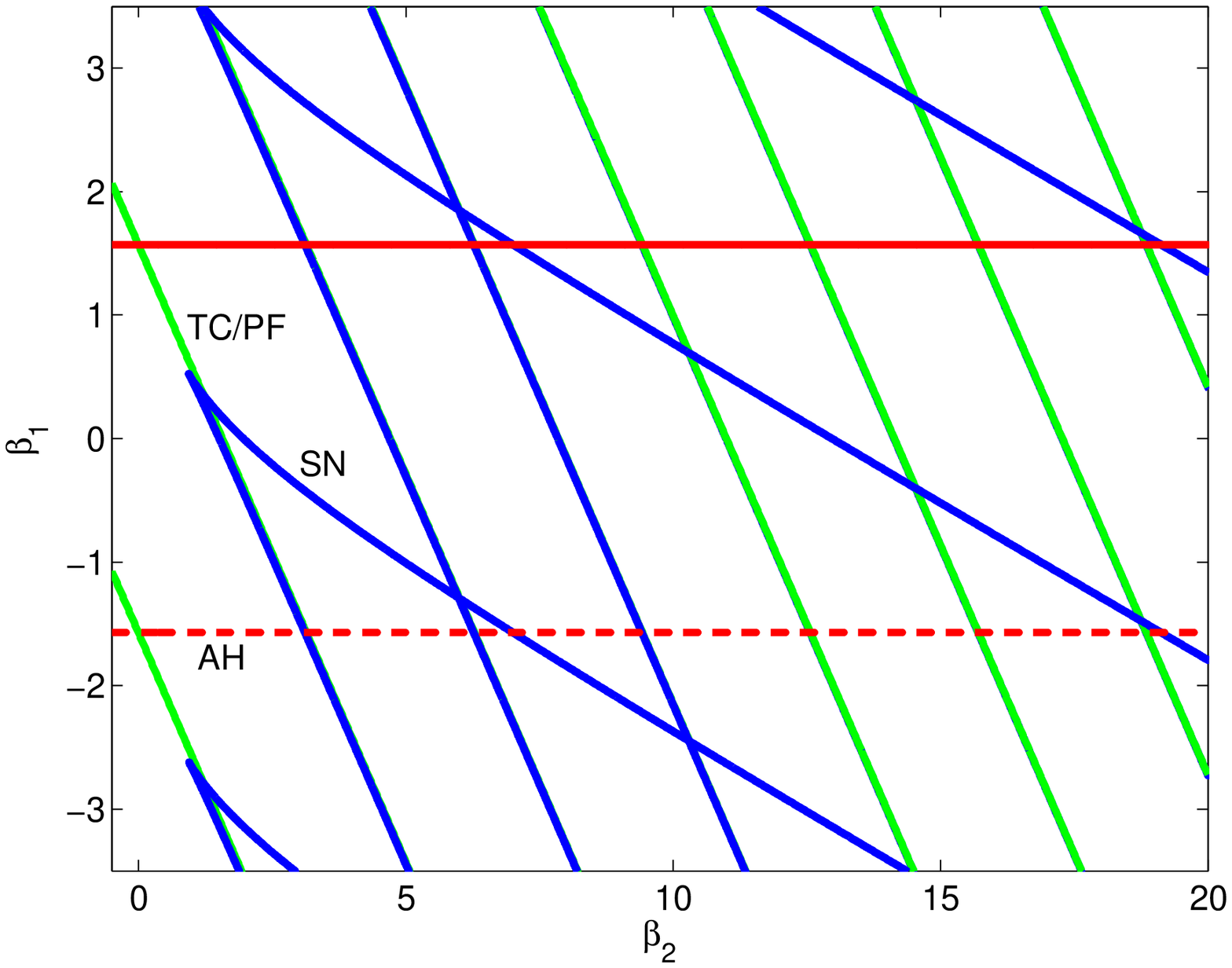}\hfill
(b)\includegraphics[width=0.46\columnwidth]{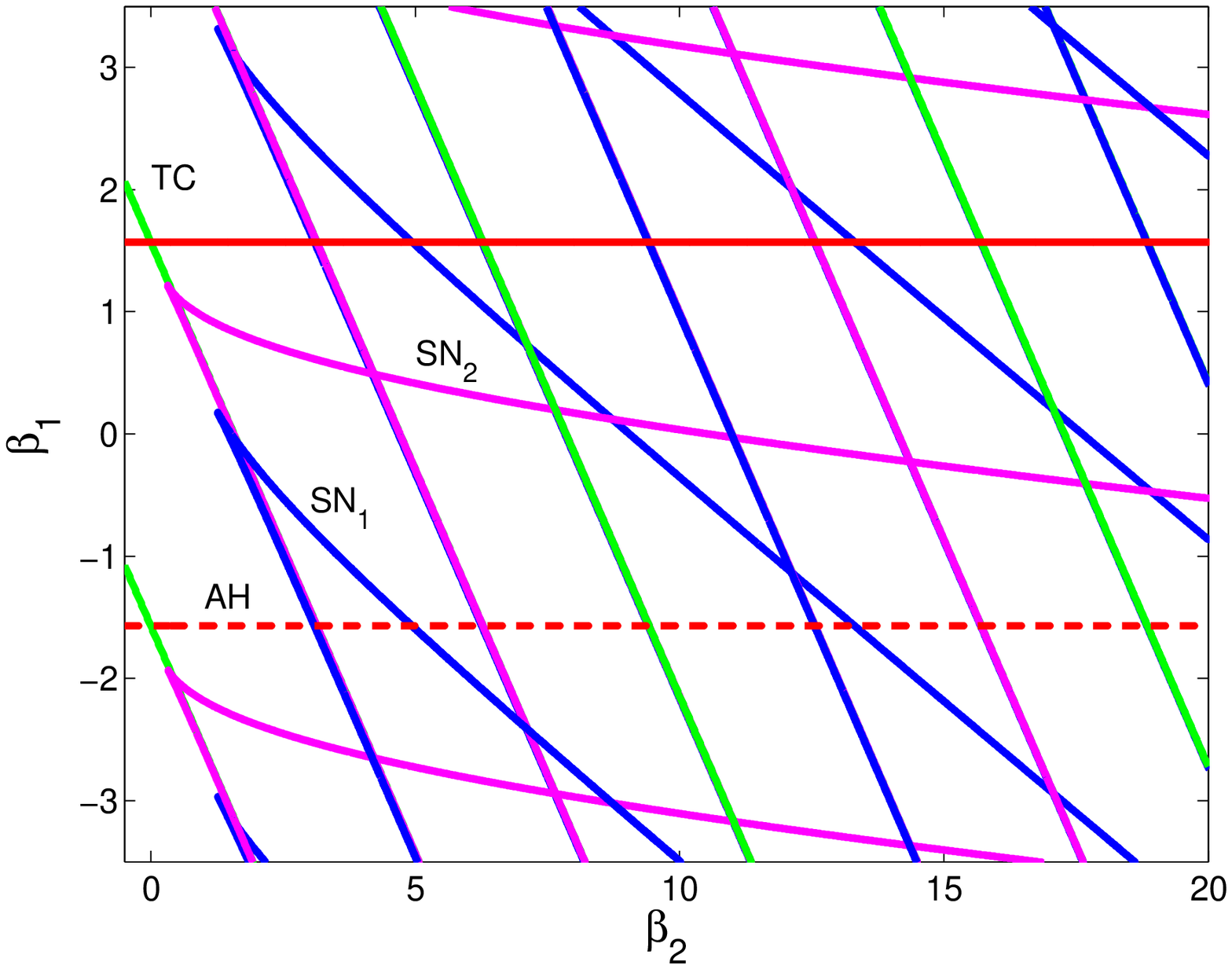}
\caption{Bifurcation diagrams for (a) $N=4$ and (b) $N=5$
oscillators
in the $(\beta_2,\beta_1)$ parametric plane. In (b): $SN_1$ --- the line of  saddle--node
bifurcation on $S_4\times S_1$ invariant line, $SN_2$ --- the same with the $S_3\times S_2$ 
invariant line. }
\label{fig:45Diagr}
\end{figure}

The bifurcation analysis of higher--dimensional cases (for $N\ge
5$) shows similar results. The bifurcation diagram consists of
three types of lines: a straight line of an Andronov--Hopf bifurcation,
a straight line of a transcritical bifurcation, and lines of saddle--node
bifurcations on invariant lines with the symmetry $S_p\times
S_{N-p}$, $p=1,\dots, N-1$.  All the saddle--node bifurcation lines
have similar ``tongue-like'' form. The
``tongue'' is formed by two border lines: one lies
left of the straight line of a transcritical bifurcation and  approaches
this line asymptotically with increasing of parameter $\beta_2$,
 the second border line crosses the TC--lines.  The saddle--node
bifurcation generates a sink and source only on the invariant line with
symmetry $S_1\times S_{N-1}$, while on other invariant lines with the symmetry
$S_p\times S_{N-p}$, $p\not =1$ a
pair of saddles appears. Therefore, there exist only
 $1\times (N-1)$ stable clusters.  Furthermore, heteroclinic cycles
and stable limit cycles appear at a saddle--node
bifurcation with $S_1\times S_{N-1}$ symmetry only (the corresponding
bifurcation lines are drawn with blue in Fig.~\ref{fig:45Diagr}).

Let us discuss the codimension-two points marked in Fig.~\ref{fig:3Diagr}. At point $C$ two
borders of the saddle-node tongue meet. Only one stable state is involved in both 
saddle-node bifurcations (say, on the left line states $1$ and $2$ are created, while on the right line
state $2$ annihilates with state $3$), at the codimension-two point $C$ all three involved steady states meet.
At another codimension-two point $A$ the type of the saddle-node bifurcation changes. On one side 
(left to the point $A$) no heteroclinic cycle appears at the saddle-node, while right to point
$A$ the transition can be saddle-node/heteroclinic, provided $-\pi/2<\beta_1(A)<\pi/2$ and the line has symmetry 
$S_1\times S_{N-1}$. We have checked that the latter condition holds for $N=3,\ldots,8$ only and
for $N\geq 9$ one has $\beta_1(A)<-\pi/2$. Thus, for $N\geq 9$ there is
no saddle-node/heteroclinic transition. Finally, the point $B$ on the bifurcation diagram Fig.~\ref{fig:3Diagr}
corresponds to a degenerate situation depicted in the middle panel of Fig.~\ref{fig:hc}.

\begin{figure}
\centering
\includegraphics[width=0.6\columnwidth]{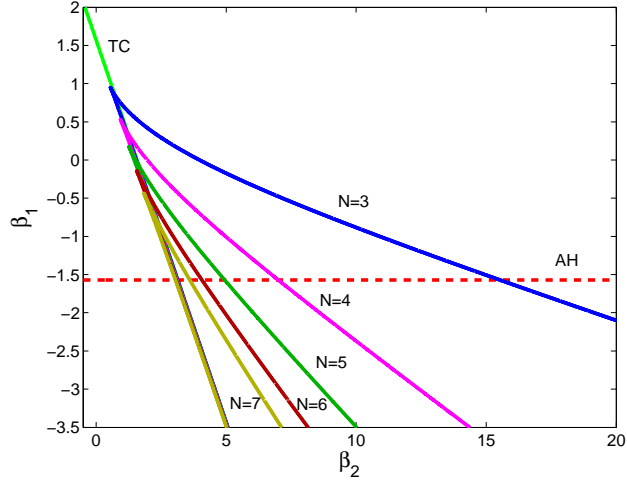}
\caption{The comparative diagram of saddle--node
bifurcation lines on the $(\beta_2,\beta_1)$ parametric plane for the cases 3, 4, 5, 6, and 7 oscillators.
The tip of the ``tongues'' shifts down as $N$ grows, and for $N\geq 9$ 
is below thee line of Andronov-Hopf bifurcation $\beta_1=-\pi/2$.}
\label{fig:Compdiag}
\end{figure}

We show bifurcation diagrams on the planes $(\beta_1,\beta_2)$ for four and five coupled 
oscillators in Fig.~\ref{fig:45Diagr}. The structure of these diagrams is basically the same as for
three oscillators 
Fig.~\ref{fig:3Diagr}, but with some quantitative changes. To clarify these changes we compare in 
Fig.~\ref{fig:Compdiag} the basic saddle-node ``tongues'' for $N=3,\ldots,7$. One can see that with 
increase  of $N$ the tip shifts down and for a fixed $\beta_1\approx 0$ the saddle-node bifurcation
can be observed for a small number of oscillators only. Thus, for a fixed $\beta_1\approx 0$ the loss
of full synchrony with increase of $\beta_2$ occurs as direct transition from full synchrony to
periodic oscillations via a transcritical bifurcation (bottom raw in Fig.~\ref{fig:hc}), and not via
clustered states.

\section{Conclusion}
In this paper we have performed a detailed bifurcation analysis of the nonlinear generalization
of the Sakaguchi-Kuramoto model of globally coupled phase oscillators. The main novelty in addition
to the 
consideration in the framework of WS theory~\cite{Pikovsky-Rosenblum-09} is the characterization
of cluster states that in terms of phase differences appear (via a transcritical, a pitchfork, or a 
saddle-node bifurcation) 
as steady states on invariant lines of the corresponding cluster configurations. At saddle-node bifurcations these
steady states disappear via heteroclinic cycles. Remarkably, heteroclinic cycles
in this model are not destroyed but remain to exist (for other examples of
heteroclinic cycles in ensembles of identical phase oscillators 
see~\cite{Hansel93,Kori-Kuramoto-01,Ashwin-etal-06}).
 This is related to the partial integrability of the system
resulting from the WS theory. According to WS, because the equations have $N-3$ constants of
motion, periodic orbits form the families of corresponding dimensions, the heteroclinic cycles form the limiting cases
of these families, describing cycles that include nearly-clustered states. 

The analysis performed in this paper complemented the conclusion on the transition from full to partial synchrony
in nonlinearly coupled oscillator ensembles, made in~\cite{Pikovsky-Rosenblum-09}. We have demonstrated that
for small ensembles the transition is of the type ``full synchrony'' $\to$ ``cluster state'' $\to$ ``periodic/quasiperiodic 
partially synchronous state'' occurs, while for a large number of oscillators a direct transition
``full synchrony'' $\to$ ``periodic/quasiperiodic 
partially synchronous state'' is typical.

\textbf{Acknowledgement} O.B. thanks DFG (Project ``Collective phenomena and 
multistability in networks of dynamical systems'') for support. 
We thank M. Rosenblum, P. Ashwin, and M. Wolfrum for useful discussions.



\end{document}